\documentclass[preprint]{revtex4}

   \def\rb{{\bf r}}      \def\Gb{{\bf G}}
         \def\Qb{{\bf Q}}
\def\ee{{\rm e}}   
\def\ii{{\rm i}}     \def\Re{{\rm Re}}  
    \def\Eb{{\bf E}}   
         
      \def\eb{{\hat{\bf e}}}
   \def\Kb{{\bf K}}      

\begin{document}

\title{Electron induced light emission in photonic crystals}

\author{L. A. Blanco$^{1,}$\footnote{phone number: ++34 943015421 \\ fax number: ++ 34 943015600 \\ e-mail address: lablancoj@sc.ehu.es}
          and F. J. Garc\'{\i}a de Abajo$^{1,2}$} 
\affiliation{             $^1$Donostia International Physics Center (DIPC), \\
             Aptdo. 1072, 20080 San Sebasti\'{a}n, Spain \\
        $^2$Centro Mixto CSIC-UPV/EHU, \\
             Aptdo. 1072, 20080 San Sebasti\'{a}n, Spain}

\date{\today}


\begin{abstract}
ID:A1857 \\ The interaction of a fast electron with a photonic crystal is
studied by solving the Maxwell equations exactly for the
external field provided by the electron in the presence of the
crystal. The polarization currents and charges produced by the passage
of the electron give rise to the emission of the so-called Smith-Purcell
radiation. The emitted light probability is obtained by integrating the Poynting
vector over planes parallel to the crystal at a large distance from the latter.
Both reflected and transmitted light components are analyzed and related to the photonic
band structure of the crystal. Emission spectra are compared with the energy
loss probability and also with the reflectance and transmittance of the
crystal. \\
Keywords: Electron solid interactions, photonic crystals, electroluminiscence.
\end{abstract}
\pacs{41.75.Ht,79.20.Kz,41.20.Bt,03.50.De,79.20.Uv}
\maketitle

\section{Introduction}
\label{SecI}

Photonic crystals have received considerable attention as many technological applications are being
proposed including control of atomic and molecular
spontaneous emission \cite{byk} or light guiding and confinement \cite{yab, joh, joa1, joa2}. 
The main physical aspect that is used for these purposes is
the existence of photonic bandgaps in such systems, that can lie in
the near infrared  \cite{ho,ebb,thi} or in the visible \cite{mor}, which allow 
light to follow a determined path depending on the structure of the crystal.
Despite recent advances in the
production of photonic crystals that operate in both regions of the light spectrum \cite{bla}, 
fabrication defects can still limit their applicability. However, a quantitative determination of the number
and type of the defects in a crystal is difficult to perform. Electrons seem to be a valid probe to 
that end.  Here, we study the interaction of a fast electron with a crystal as a first approach to
this problem.

In particular, a fast electron moving near a photonic crystal can polarize its constituents inducing
charges and currents that cannot follow the electron motion and suffer acceleration as they evolve 
in the inhomogeneous crystal, giving rise to light emission. This is similar to the
so-called Smith-Purcell effect \cite{fra,smi,bac,hae,pen,jgasmi,oht}.

The present work is intended to provide a new way to characterize
photonic crystals by observing the spectrum of the light emitted 
when electrons are 
travelling parallel to the surface of the crystal. The
electric field of the moving electron is decomposed into evanescent
plane waves \cite{tor}, and the outgoing light is
produced by the diffraction of these
waves in the crystal. This procedure is sketched in Section II.
Numerical results for a complete-band-gap photonic crystal
formed by air spheres in  silicon (inverted
opal) are shown in Section III.
We use electrons travelling within air and silicon,
in order to reflect different electrodynamic effects, 
although a possible experimental observation could be more difficult
in the latter case.
Gaussian atomic units will be used from now on,
unless otherwise specified.

\section{Theoretical framework}
\label{SecII}

A theoretical description of the interaction between an electron
and a crystal is given next.
The crystal will be composed of several layers
perpendicular to the $z$ direction, beginning in $z=0$ and
extending towards the negative-$z$ region. Outside
the crystal a medium with a dielectric function $\epsilon$
will also be considered. All media will be assumed to be non-magnetic
($\mu=1$).

The electron will be considered to follow a trajectory
described by
$\rb_t=(v t, y_0, z_0)$, with $z_0>0$, and its
electric field is found to be, in frequency space $\omega$ and
in the absence of the crystal,
   \begin{eqnarray}
      \Eb_0(\rb,\omega)=\int dQ_y \,
                        \ee^{\ii \Kb^\pm_\Qb\cdot[\rb-(0,y_0,z_0)])}
                        \Eb^\pm_\Qb,
   \label{e2}
   \end{eqnarray}
where 
$\Kb^\pm_\Qb = (\Qb , \pm \ii \Gamma_\Qb)$, $\Qb = (\omega/v,Q_y)$
is the two-dimensional momentum parallel to the surface, and
$\Gamma_\Qb^2 = Q^2 - \omega^2 \epsilon/c^2$, with
$\Re \{ \Gamma_\Qb \} > 0$. In this expression,  
$ \Eb^\pm_\Qb\,\exp(\ii \Kb^\pm_\Qb\cdot\rb) $
is a plane wave, which  
can be expressed as a sum of $s$ and $p$ components, 
$      \Eb^\pm_\Qb=E^\pm_{\Qb,s}\,\eb_{\Qb,s}^\pm + E^\pm_{\Qb,p}\,\eb_{\Qb,p}^\pm $.
The $+$ ($-$) sign in these expressions stands for a
wave moving towards the positive- (negative-)$z$ region. 
When the electron moves in vacuum, all of these waves 
are evanescent ($\Gamma_\Qb$ is real and positive).
However, when it moves in a medium described by $\epsilon>0$ and $v\sqrt{\epsilon} > c$, 
some of those waves represent propagating Cherenkov radiation.

Due to the crystal symmetry,
an incident wave with momentum
$\Kb^-_\Qb$ will only produce a discrete set of reflected (transmitted) waves of
momentum $\Kb^+_{\Qb+\Gb}$
($\Kb^-_{\Qb+\Gb}$), $\Gb$ being a reciprocal
surface lattice vector. 
Therefore, 
the reflected and transmitted wave amplitudes can be expressed as
   \begin{eqnarray}
     \nonumber
      [E^+_{\Qb+\Gb,\sigma}]^r & = & \sum_{\sigma'}
                             R_{\Qb\Gb}^{\sigma\sigma'}
                             [E^-_{\Qb,\sigma'}]^i, \\
      \label{e1}
      [E^-_{\Qb+\Gb,\sigma}]^t & = & \sum_{\sigma'}
                             T_{\Qb\Gb}^{\sigma\sigma'}
                             [E^-_{\Qb,\sigma'}]^i, 
   \end{eqnarray}
where $\sigma$ and $\sigma'$ run over polarizations $s$
and $p$, the super-indices $r$, $t$ and $i$ stand for reflected,
transmitted  and
incident components, respectively, and
$R_{\Qb\Gb}^{\sigma\sigma'}$ and
$T_{\Qb\Gb}^{\sigma\sigma'}$ are the reflection and transmission coefficients
of the crystal.
These coefficients are calculated using 
the layer KKR method \cite{ste}, in which the self-consistent electric
field is constructed in terms of multipole expansions around the crystal objects
by multiple scattering within each layer and later among layers.

Each of the incident plane waves in Eq. (\ref{e2})
gives rise to a set of reflected and transmitted waves whose
amplitudes are obtained according to Eq.\ (\ref{e1}). The
electric field in the positive-$z$ region can be constructed as
the sum of $\Eb_0$ and the reflected field,
   \begin{eqnarray}
      \Eb(\rb,\omega)=\Eb_0(\rb,\omega)+\Eb_r(\rb,\omega),
   \nonumber 
   \end{eqnarray}
where
   \begin{eqnarray}
      \Eb_r(\rb,\omega)=\sum_{\Gb,\sigma\sigma'} \int dQ_y \,\,
          \ee^{\ii \Kb^+_{\Qb+\Gb}\cdot\rb} \,
          R_{\Qb\Gb}^{\sigma\sigma'} \, [E^-_{\Qb,\sigma'}]^i \,\,
          \eb_{\Qb+\Gb,\sigma}^+.
   \nonumber 
   \end{eqnarray}
A similar expression is found for the transmitted electric field
at the other side of the crystal (negative $z$'s):
    \begin{eqnarray}
	\Eb_t(\rb,\omega)=\sum_{\Gb,\sigma\sigma'} \int dQ_y \,\, 
          \ee^{\ii \Kb^-_{\Qb+\Gb}\cdot\rb} \,
          T_{\Qb\Gb}^{\sigma\sigma'} \, [E^-_{\Qb,\sigma'}]^i \,\,
          \eb_{\Qb+\Gb,\sigma}^-.
   \nonumber 
   \end{eqnarray}

The light emission probability per unit length is calculated by means of
the projection of the Poynting vector ${\cal P}$ over the $z$ and $-z$
directions (for reflected and transmitted components, respectively), which
is integrated over the time and over a plane parallel to the crystal
surface. The resulting expression can be written

\begin{equation}
\int dt \int dx \int dy {\cal P} \cdot (\pm \hat{z})= \int_0^\infty \omega d\omega P(\omega),
\end{equation}

where $P(\omega)$ can be interpreted as the emission probability per unit
length and per energy range $\omega$. The average over parallel impact
parameters $y_0$ has also been performed. When the Cherenkov condition
$v^2 \epsilon > c^2$ is not fulfilled, the direct waves of
Eq. (\ref{e2}) are all evanescent, and the light emission by reflection and
transmission at the crystal (i.e. emission towards the $z>0$ and $z<0$
regions, respectively denoted reflected emission
and transmitted emission hereafter) reduces to

\begin{eqnarray}
    P_r(\omega) & = & \frac{1}{2 \pi  k^2} \sum_{\Gb \sigma} \, ' \int dQ_y
\vert r_{\Qb \Gb \sigma} \vert^2 \Delta_{\Qb+\Gb} e^{-2 \Delta_\Qb z_0} ,
\nonumber \\
    P_t(\omega) & = & \frac{1}{2 \pi  k^2} \sum_{\Gb \sigma} \, ' \int dQ_y
\vert t_{\Qb \Gb \sigma} \vert^2 \Delta_{\Qb+\Gb} e^{-2 \Delta_\Qb z_0} ,
\nonumber
\end{eqnarray}

where 

\begin{eqnarray}
     \nonumber
      r_{\Qb \Gb \sigma} & = & \frac{\ii Q_y \omega }{Q \Gamma_\Qb c^2} R_{\Qb \Gb}^{\sigma s} -
      \frac{k}{vQ\sqrt{\epsilon}} R_{\Qb \Gb}^{\sigma p}, \\
     \nonumber
      t_{\Qb \Gb \sigma} & = & \frac{\ii Q_y \omega }{Q \Gamma_\Qb c^2} T_{\Qb \Gb}^{\sigma s} -
      \frac{k}{vQ\sqrt{\epsilon}} T_{\Qb \Gb}^{\sigma p},  \\
     \nonumber
      \Delta_\Qb & = & \vert \Gamma_\Qb \vert
\end{eqnarray}
and the prime in the summation over $\Gb$ indicates that only non-evanescent outgoing waves
($\Gamma_{\Qb+\Gb}^2 <0$)
must be included.

On the other hand, if $v^2 \epsilon > c^2$, there exists an
interference between the direct Cherenkov radiation and the reflected field. The 
emission probability expressions take a more involved form in this case:

\begin{eqnarray}
    \nonumber P_r(\omega) & = & \frac{1}{2\pi k^2} \sum_\sigma \left\{ \int_{- \infty}^{-Q_0} dQ_y \sum_\Gb \, ' 
      \vert r_{\Qb \Gb \sigma} \vert^2 \Delta_{\Qb+\Gb} e^{-2 \Delta_\Qb z_0} \right. + \\
    \nonumber & + & \int_{-Q_0}^{Q_0} dQ_y \left[ \Delta_\Qb \vert f_\sigma \vert^2 +
       2 \Re \left( \Delta_\Qb f_\sigma r_{\Qb 0 \sigma} e^{-2 \ii \Gamma_\Qb z_0} \right) +
       \sum_\Gb \, ' \vert r_{\Qb \Gb \sigma} \vert^2 \Delta_{\Qb+\Gb} \right] + \\
    \nonumber & + & 
                     \left. \int_{Q_0}^{\infty} dQ_y \sum_\Gb \, ' 
      \vert r_{\Qb \Gb \sigma} \vert^2 \Delta_{\Qb+\Gb} e^{-2 \Delta_\Qb z_0} \right\}, \\
    \nonumber P_t(\omega) & = & \frac{1}{2\pi k^2} \sum_{\sigma \Gb} \, ' \left\{ \int_{- \infty}^{-Q_0} dQ_y  
      \vert t_{\Qb \Gb \sigma} \vert^2 \Delta_{\Qb+\Gb} e^{-2 \Delta_\Qb z_0} \right. + \\
    \nonumber
    & + & \int_{- Q_0}^{Q_0} dQ_y  
      \vert t_{\Qb \Gb \sigma} \vert^2 \Delta_{\Qb+\Gb} + \\
     \nonumber & + & \left. \int_{Q_0}^{\infty} dQ_y  
      \vert t_{\Qb \Gb \sigma} \vert^2 \Delta_{\Qb+\Gb} e^{-2 \Delta_\Qb z_0} \right\} ,
\end{eqnarray}
where $Q_0^2=k^2\epsilon - \omega^2/v^2$,  
$f_p =  - k/(vQ\sqrt{\epsilon})$ and 
$f_s = \ii Q_y \omega /(Q \Gamma_\Qb c^2)$.
The integrands in these expressions describe the reflected and transmitted wave components
for external evanescent ($\vert Q_y \vert > Q_0$) and propagating
($\vert Q_y \vert < Q_0$) waves. In particular, the first term inside the square
bracket of $P_r$ reduces to the Cherenkov emission probability in a bulk homogeneous medium,
while the remaining terms describe reflected and interference components.

\section{Results and discussion}
\label{secIII}

We have applied the previous formalism to the case of a 100-keV 
electron moving parallel to a slab consisting on several fcc(100) and fcc(111) planes of
air spheres in silicon. The gap characteristics of this system have
been theoretically described in \cite{bu}: a complete
photonic bandgap is opened when the ratio between the sphere radius $r$ and
the cubic lattice constant $a$ lies in
between 0.335 and 0.374. 
We have chosen an intermediate value of $r/a=0.342$, and  the
frequency range is taken close to the communications wavelength, $\lambda=1.55 \mu$m, in
which case $\epsilon_{\rm Si}=11.9$.

Results for crystals composed of 8 fcc(111) planes are shown in
what follows, although we have also carried out calculations for
8 fcc(100) planes and the results are qualitatively similar.
Two different
types of calculations are presented. a) In Fig. 1, the crystal is embedded in Si,
so that the electron moves within silicon outside
the crystal. In this case, the electron produces Cherenkov radiation even without the
presence of the crystal whenever the electron velocity is larger than the speed of light in
the medium, which is the case at 100 keV. b)
In Fig. 2, the crystal is embedded in air and the electron moves also
outside the crystal, that is in air. Therefore, no direct Cherenkov radiation is produced.

As the total energy must be conserved in the present case in which the material is transparent (that is,
$\epsilon$ is real) the sum of the reflected and transmitted probabilities must be the same
as the electron energy loss probability, which has been calculated 
from the retarding force exerted on the electron by the induced electric field, as
discussed elsewhere \cite{gar, eels}.

The main difference between the two cases considered in Figs. 1 (a) and 2 is that when Cherenkov radiation 
is emitted 
the energy loss and the reflected emission are very much enhanced [Fig. 1 (a)].
This is due to the fact that the Cherenkov
emission is dominant as compared to the diffraction of evanescent waves.
For the transmitted
light, as expected, the emission probability within the gap frequencies is
strongly suppressed in both cases.

When the electron travels in air (Fig. 2), the energy loss 
and the reflected light probabilities are much smaller than in the previous case.
However, the order of magnitude of the
transmitted field is the same as in the Cherenkov case. Notice that huge peaks are
produced in the reflection curve near the bandgap region, probably
connected with band edge effects.
This type of energy losses are well below the current energy resolution of STEM machines (around 0.5 eV), but
since the sample is transparent, any energy loss must be converted into photon emission, for which
the resolution is much better, as shown in \cite{yam}, and sufficient to resolve the features
discussed in this work.
The integrated area of the sharp peaks in Fig. 2
amounts to approximately $5 \times 10^{-6}$ photons per electron per nm, which for
path lengths of the order of a several microns, would result in a
measurable photon intensity of a percentage of the number of electrons.

In both cases the gap regions agree reasonably well with those found for the
transmission of external electromagnetic
waves under normal and Cherenkov-angle incidence conditions  as well
as with the band structure of the system, which are shown for
the first case (electron moving in Si) in Figs. 1 (b)-(c).
Note that the transmitted emission gap [Fig. 1 (a)] is
wider than the complete gap (vertical dashed lines throughout all the
Fig. 1), since the latter corresponds to a particular selection of
incident directions (the Cherenkov cone). The same is true
for the transmittance [Fig. 1(b)]; actually, the case of normal incidence
corresponds to the $\Gamma$ point.
The depletion of light transmission within the gap could
be used to detect defaults (stacking faults, missing atoms, etc.), whose
relatively small contribution to the emission would be amplified because
that would be the primary origin of light transmitted within the gap.

Finally, in Fig. 3 we show the dependence of the 
light emission probability on the distance between
the electron and the crystal, for a given frequency within the bandgap. As 
expected, both the transmitted and reflected probabilities decrease
when increasing the electron-crystal separation, but
it is
interesting how the magnitude of the interference between the direct (Cherenkov)
and the reflected fields produces an oscillatory pattern at the front
side of the crystal. For the transmitted field, the probability reaches an asymptotic value which
comes from the transmission of only
Cherenkov incident waves.

\acknowledgments

The authors gratefully acknowledge help and support from the
Spanish Ministerio de Ciencia y Tecnolog\'{\i}a and the Basque
Government.

\newpage

FIG. 1: (a) Frequency variation of the energy loss probability (thick solid curve),
the reflected plus direct light emission probability (dashed curve) and the transmitted
light emission probability (thin solid curve) for a 100-keV electron moving in front of
8 fcc(111) planes of air spherical voids of radius $r$=418 nm in silicon
($\epsilon=11.9)$,
where the lattice constant $a$ is 1222 nm (the filling fraction
is 67\% and the separation between sphere centers is $d$=873 nm).
 The electron is moving inside Si
at a distance of 876 nm from the void spheres.
(b) Transmittance for light inciding normally (thick curve) and with the
Cherenkov angle $\theta_c=68.1^\circ$ (thin curve) for the same system.
(c) Bulk band structure projected on the (111) plane
for the same system, where the shaded zones
indicate regions where no propagating electromagnetic waves are allowed.
A full gap is observed in the range $\omega d/c \approx 3.23-3.35$.

FIG. 2: Frequency variation of the energy loss probability (thick solid curve),
the reflected plus direct light emission probability (dashed curve) and the transmitted
light emission probability (thin solid curve) for a 100-keV electron moving in front of
8 fcc(111) planes of air spherical voids of radius $r$=418 nm in silicon
($\epsilon=11.9)$,
where the lattice constant $a$ is 1222 nm (the filling fraction
is 67\% and the separation between sphere centers is $d$=873 nm).
The surface of the crystal
is 131 nm from the last sphere. The electron is moving in air
at a distance of 800 nm from this surface.

FIG. 3: Variation of the reflected plus direct 
emission probability (dashed curve) and the transmitted emission probability
(solid curve)
with the distance between a 100-keV electron and the first of the
8 fcc(111) planes of air spherical voids of radius
$r$=418 nm in silicon, where the lattice constant $a$ is 1222 nm.
The emitted light wavelength is $\lambda= 1.68 \mu$m (i.e,
$\omega d/c=3.25$). The asterisk ($\ast$) corresponds to the 
electron-spheres distance
of Fig. 1 (a).
\end{document}